\documentclass[aps,prb,reprint,superscriptaddress,floatfix]{revtex4-2}
\usepackage{comment}
\usepackage{bm}
\usepackage[final]{graphicx}
\usepackage{epsfig}
\usepackage{color}
\usepackage{verbatim}
\usepackage[OT1]{fontenc} 
\usepackage{dcolumn}
\usepackage{xcolor}
\usepackage[final]{graphicx}
\usepackage{ragged2e} % for the \justifying macro
\usepackage{amsmath}
\usepackage{bm}
\usepackage{floatrow}
\usepackage{physics}
\usepackage{times}
\usepackage[colorlinks=true,allcolors=blue]{hyperref}%
\usepackage{relsize}
\usepackage[capitalise]{cleveref}

\usepackage{lineno} % This can be removed in the future
\usepackage{tikz}
\usetikzlibrary{shapes.geometric}
\usetikzlibrary{decorations.markings}

\definecolor{DarkBlue}{rgb}{0.1,0.1,0.5}
\definecolor{Red}{rgb}{0.9,0.0,0.1}
\definecolor{Green}{rgb}{0.0,0.99,0.0}

\newcommand{\figref}[2]{\hyperref[#1]{~\ref{#1}{#2}}}

\begin{document}
%%%%%% Title %%%%%%
\title{Detection of chiral spin fluctuations driven by frustration in Mott insulators}
%%%%%% Authors %%%%%%
\author{Kuan H. Hsu}
\affiliation{Department of Materials Science and Engineering, Stanford University, CA 94305, USA}
\affiliation{Stanford Institute for Materials and Energy Sciences,
SLAC National Accelerator Laboratory, 2575 Sand Hill Road, Menlo
Park, CA 94025, USA}

\author{Chunjing Jia}
\affiliation{Department of Physics, University of Florida, FL 32611, USA}

\author{Emily Z. Zhang}
\affiliation{Stanford Institute for Materials and Energy Sciences,
SLAC National Accelerator Laboratory, 2575 Sand Hill Road, Menlo
Park, CA 94025, USA}
\affiliation{Geballe Laboratory for Advanced
Materials, Stanford University, CA 94305, USA}

\author{Daniel Jost}
\affiliation{Stanford Institute for Materials and Energy Sciences,
SLAC National Accelerator Laboratory, 2575 Sand Hill Road, Menlo
Park, CA 94025, USA}

\author{Brian Moritz}
\affiliation{Stanford Institute for Materials and Energy Sciences,
SLAC National Accelerator Laboratory, 2575 Sand Hill Road, Menlo
Park, CA 94025, USA}

\author{Rudi Hackl}
\affiliation{School of Natural Sciences, Technische Universit{\"a}t M{\"u}nchen, Garching 85748, Germany}
\affiliation{IFW Dresden, Helmholtzstrasse 20, Dresden 01069, Germany}

\author{Thomas P. Devereaux}
\email{tpd@stanford.edu}
\affiliation{Department of Materials Science and Engineering, Stanford University, CA 94305, USA}
\affiliation{Stanford Institute for Materials and Energy Sciences,
SLAC National Accelerator Laboratory, 2575 Sand Hill Road, Menlo
Park, CA 94025, USA}
\affiliation{Geballe Laboratory for Advanced
Materials, Stanford University, CA 94305, USA}
\date{\today}

\begin{abstract}

Topologically ordered states, such as chiral spin liquids, have been proposed as candidates that host fractionalized excitations. However, detecting chiral character or proximity to these non-trivial states remains a challenge. Resonant Raman scattering can be a powerful tool for
detecting chiral fluctuations, as the $A_{2g}$ channel probes excitations with broken time-reversal symmetry and local chiral order. Here, we use exact diagonalization to characterize the resonant $A_{2g}$ channel, alongside two-magnon scattering in $B_{1g}$ and $E_g$ channels, for the Hubbard model on lattices with increasing levels of geometric spin frustration, where tuning the incident energy near the Mott gap reveals strong chiral spin excitation intensity. Increased spin frustration in the Mott insulator results in an overall softening of the Raman $A_{2g}$ response, indicating a tendency toward low energy chiral-chiral fluctuations in Mott insulators with magnetic frustration and proximity to chiral spin liquid states that can potentially be tuned by external perturbations.
\end{abstract}
\maketitle

\section{Introduction} 

In strongly correlated systems, frustration can give rise to exotic phases such as quantum spin liquids (QSL), a highly entangled state that lacks magnetic ordering even down to zero temperature~\cite{balents2010spin,broholm2020quantum}. A sub-class of QSLs, known as chiral spin liquids (CSL), can host fractionalized quasiparticle excitations with non-Abelian statistics, giving rise to unusual phenomena such as the fractional quantum Hall effect~\cite{kalmeyer1989theory}, and unconventional superconductivity~\cite{laughlin1990properties,Wen1989Chiral}. These states, which break both parity and time-reversal symmetry, may occur even in systems whose parent Hamiltonians are time-reversal invariant (Kalmeyer-Laughlin type). For example, theoretical and numerical studies suggest that such CSLs may be realized in geometrically frustrated Hubbard-type models.~\cite{Yang1993possibleSL,Yao2007ExactCSL,Bauer2014kagomeCSL,Messio2017CSLDM,gong2019csl,szasz2020chiral}. 

Experimentally detecting CSLs remains challenging, as there are no direct probes for their fractionalized excitations. However, one can measure the scalar spin chirality (SSC), offering an avenue for detecting unusual spin textures and time-reversal symmetry breaking fluctuations in frustrated magnets. For instance, a non-vanishing SSC in a spin liquid phase can be a useful indicator to determine whether or not a QSL is chiral. In magnetically ordered ground states from $SU(2)$ symmetric systems, signatures of low lying excited states with chiral character can also serve as an indication that a CSL state may lie in close proximity of the ground state. Such states could potentially be stabilized with external perturbations. Nevertheless, detecting SSC both experimentally and theoretically have been less explored as a tool for identifying chiral character in putative QSL states. 

To date, several experimental techniques for detecting SSC in magnetic systems have been proposed. Shastry and Shraiman first pointed out that fluctuations of the SSC operator couple to the $A_{2g}$ channel in Raman scattering on a square lattice geometry~\cite{ShastryRaman,ShastryRaman_long}. Resonant inelastic scattering (RIXS) in the pre-edge region, with a dipole excitation of a core electron to an off-site valence orbital, has also been proposed as a means to measure SSC \cite{ko2011proposal}; however, the use of high-energy x-rays limits the experimental resolution. Neutron scattering also may be able to detect SSC if the chiral spin fluctuations couple to $S_z$ fluctuations~\cite{lee2013proposal}. 

Even with these proposals, Raman spectroscopy remains the only realized experimental technique that directly detects chiral spin fluctuations.  As an optical probe, Raman scattering also allows for smaller sample sizes, making it easily accessible for single-crystal and thin-film experiments~\cite{wulferding2019raman}. Although limited to providing information for very small momentum transfer, Raman scattering serves as a well-established technique for analyzing dynamical properties by selectively utilizing photon polarization to project out the symmetry group of excited states. Measurements of the $A_{2g}$ channel have been made in insulating cuprates~\cite{SulewskiA2g1991,liu1993novel,chelwani2018magnetic}, a kagome spin liquid candidate~\cite{wulferding2010interplay}, and a heavy fermion superconductor~\cite{kung2015chirality}.

In this paper, we study the Raman scattering cross-section of the two-dimensional Hubbard model at half-filling. We calculate the resonant Raman cross-section directly by using exact diagonalization (ED) and perturbation theory for the light-matter interaction (Fermi's Golden Rule)~\cite{Tom2007Review}. Specifically, we focus on Raman scattering in the $A_{2g}$ channel, which probes excited states that break time-reversal symmetry. The dynamical properties of the Hubbard model are modified by tuning the frustration with different lattice geometries, with increasing spin frustration from the square and honeycomb lattices to the triangular and kagome lattices. Frustration is also introduced by decreasing the strength of Coulomb interaction strength in the Hubbard model, rather than relying exclusively on a down-folded, spin-only Heisenberg picture. 

We first study the static spin correlations, showing that the static spin structure factor becomes increasingly incoherent with greater geometric frustration and weaker electron correlation. We derive and calculate the Raman $A_{2g}$ response in the Heisenberg limit as a chiral-chiral correlator, which smoothly connects to the Raman $A_{2g}$ cross-section calculated for the Hubbard model in the off-resonance regime. We observe a softening of the Raman response in both the two-magnon and the $A_{2g}$ scattering channel when the ground state appears to be in close proximity to a nascent chiral spin state for highly frustrated lattice geometries, which may be stabilized by external fields~\cite{Claassen2017dynamical}. When the incident energy is tuned near resonance, we show that the Raman $A_{2g}$ cross section is significantly enhanced and can serve as a reliable tool to detect chiral character in potential CSL candidates.

\section{Results}

\subsection{Spin structure factor} \label{sec:Structure-Factor}

\begin{figure} [!bt]
\begin{center}
\includegraphics[width=1\columnwidth]{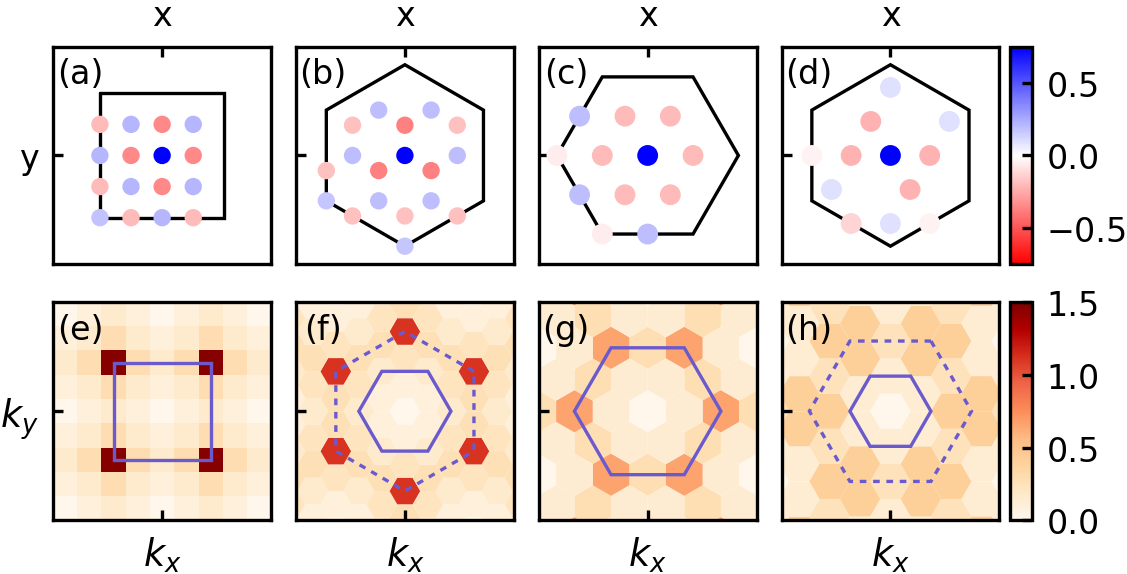}
\end{center}
\caption{\textbf{Real space spin-spin correlation $\mathbf{S_r}$ (top row) and static spin structure factor $\mathbf{S^{zz}_q}$ (bottom row) in the strong coupling limit}: Unit cells in the first row for the \textbf{(a)} $\mathbf{4 \times 4}$ \textbf{square lattice}, \textbf{(b)} $\mathbf{3 \times 3}$ \textbf{honeycomb lattice}, \textbf{(c) 12 site} (12C~\cite{Betts_cluster}) \textbf{triangular lattice}, and \textbf{(d)} $\mathbf{2 \times 2}$ \textbf{kagome lattice} are shown with the real space spin-spin correlation $S_r$ plotted as colored dots for points inside the unit cell. The corresponding first (extended) Brillouin zones are plotted in the second row for each lattice type using solid (dashed) blue lines. The intensity of the static spin structure factor $S^{zz}_q$ is plotted for the accessible $k$-points in each panel. The spin structure factor calculated using either the Hubbard model with $U=40t$ or the Heisenberg model provide consistent results.}
    \label{fig:SS_corr}
\end{figure}

We first consider the spin structure factor $S_q$ for the Hubbard model on four lattices in Fig.~\figref{fig:SS_corr}. The real space spin-spin correlations in $S_r$ for the square lattice show that spins are aligned antiferromagnetically, with the corresponding momentum-space $S_q$ strongly peaked at the M-point [($\pi$,$\pi$)] in the Brillouin zone (see Figs.~\figref{fig:SS_corr}{(a) and (e)}). Here, antiferromagnetism persists for all values of $U$.

Similar to the square lattice, the honeycomb lattice is bipartite and possesses an antiferromagnetic ground state at large enough $U$. Here, we evaluate the Hubbard model on a $3\times3$ honeycomb lattice (18 total sites). In the strong coupling limit, the spins are aligned antiferromagnetically with $S_q$ peaked at the K-point of the extended Brillouin zone (see Figs.~\figref{fig:SS_corr}{(b) and (f)}). The strength of $S_q$ gradually decreases with decreasing $U$, but the antiferromagnetic ordering persists until $U=8t$ with no change in ground state characteristics.

Geometric frustration becomes significant on triangular lattices, where spin liquid phases have been proposed to be potential ground state candidates for frustrated magnets~\cite{Shen2016,zhou2011spin} and organic Mott insulators~\cite{shimizu2003spin,kanoda2011mott}. For a triangular lattice Hubbard model in the strong coupling limit, frustration leads to a 120$^\circ$ N\'{e}el state with long-range order and $S_q$ peaked at the K-point of the Brillouin zone (see Figs.~\figref{fig:SS_corr}{(c) and (g)}), albeit less strongly than that of the honeycomb lattice. 

\begin{figure} [!bt]
\begin{center}
\includegraphics[width=1\columnwidth]{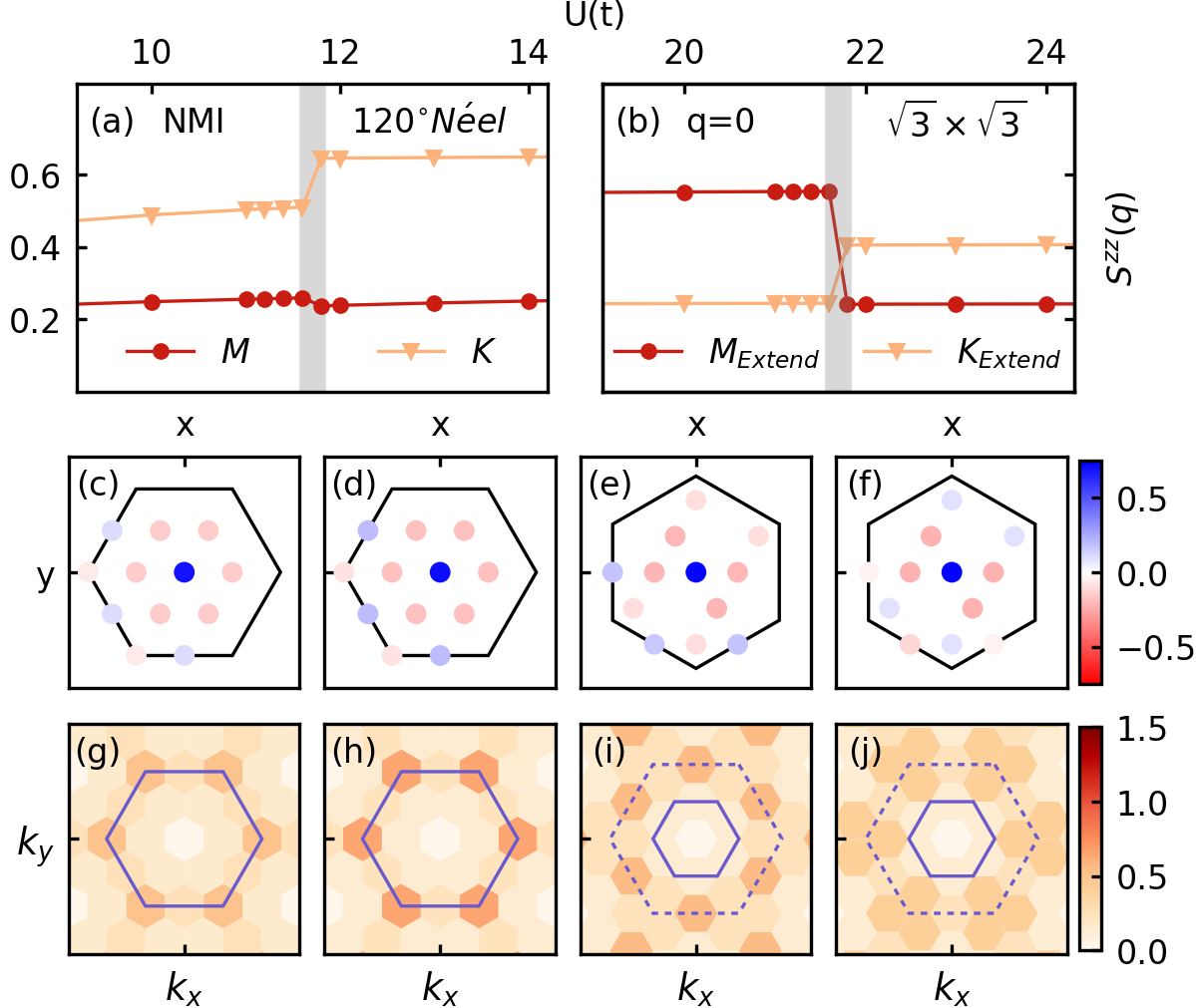}
\end{center}
\caption{\textbf{Changes in the ground states on the (a) triangular and (b) kagome lattice, characterized by real space spin-spin correlation $\mathbf{S_r}$, and static spin structure factor $\mathbf{S^{zz}_q}$}: Sharp changes in the static spin structure factor at the K and M points in the (extended for kagome lattice) Brillouin zone near ground state transitions are shown across various values of $U$. The shaded area specifies the region where the ground state transition occurs. Real-space spin-spin correlations are displayed for (c) the triangular lattice in the NMI state at $U = 11t$, and (d) the 120$^\circ$ N\'{e}el  state at $U = 12t$, with their respective static spin structure factors in panels (g) and (h). For the kagome lattice, real-space spin-spin correlations for the $q = 0$ state at $U = 21t$ and the $\sqrt{3} \times \sqrt{3}$ state at $U = 22t$ are shown in (e) and (f), with corresponding static spin structure factors in (i) and (j).}
    \label{fig:GS_change}
\end{figure}

Stronger charge fluctuations in the intermediate coupling regime can lead to an instability in the ground state where the long-range antiferromagnetic order disappears. At $U<11.8t$, $S_q$ at the K-point [($\frac{4}{3}\pi$,0)] shows a sharp decrease (Fig.\figref{fig:GS_change}{(a)}), signifying the destruction of 120$^\circ$ N\'{e}el order. Here, translational symmetry is broken and the ground state momentum shifts to the M-point, where the ground state becomes non-magnetically insulating (NMI). Here, the change from 120$^\circ$ N\'{e}el order to NMI is subtle and may be hard to observe without detailed inspection of the spin structure factor. Studies on large clusters have also indicated a ground state transition to a spin liquid candidate NMI state at similar values of $U$~\cite{shirakawa2017ground,szasz2020chiral,chen2022quantum,xu2024hubbard}. 

Electrons arranged on a kagome lattice experience the most geometric frustration. The kagome lattice consists of corner-sharing triangles where the spins on an individual plaquettes are geometrically frustrated. At large $U$ for the Hubbard model on a 2$\times$2 kagome lattice (12 total sites), $S_q$ possesses weak peaks near the K-points of the extended Brillouin zone (see Figs.~\figref{fig:SS_corr}{(d) and (h)}). At intermediate coupling where $U<21.8t$, $S_q$ is more strongly peaked at the M-points (Figs.~\figref{fig:GS_change}{(e) and (i)}). This change in the spin correlations signifies a change of spin ordering at the ground state transition, going from a $\mathrm{\sqrt{3}\times\sqrt{3}}$ ordered state to a $q=0$ ordered state \cite{messio2011lattice}. We note that a study with larger clusters also demonstrated a similar ground state transition behavior on a 2$\times$L kagome cylinder~\cite{sun2021metal}.

\subsection{Resonant Raman Scattering in the Off-Resonance Regime} \label{sec:OffRes-Raman}

\begin{figure} [!tb]
\begin{center}
\includegraphics[width=1\columnwidth]{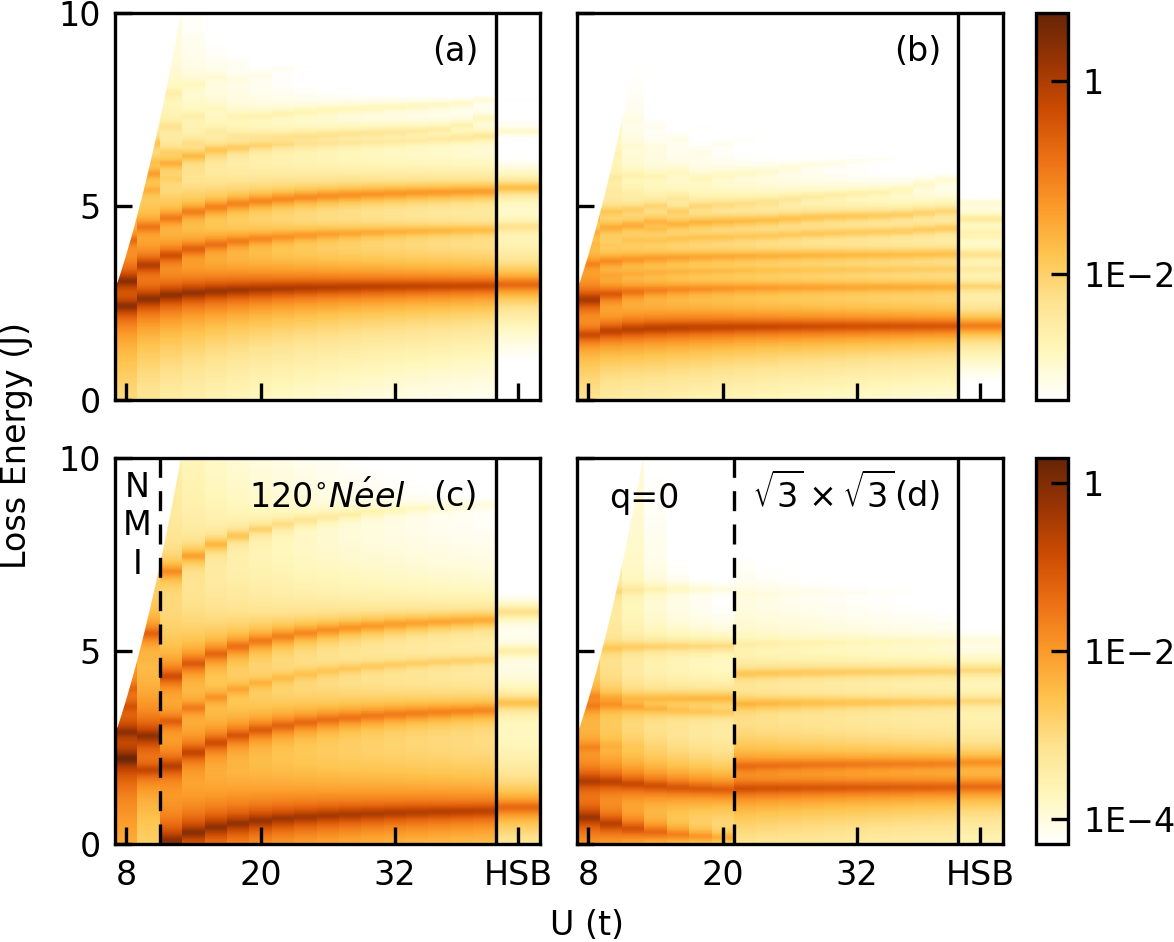}
\end{center}
\caption{\textbf{Off-resonance Raman spectra for  (a)
$\mathbf{B_{1g}}$ square lattice (b) $\mathbf{E_{g}}$ honeycomb lattice (c) $\mathbf{E_{g}}$ triangular lattice and (d)
$\mathbf{E_{g}}$ kagome lattices, with incident energy at $\mathbf{\omega_i = U/4}$}:  The spectra are calculated as a function of $U$ in the Hubbard model, while the right-most column in each panel with the label \textit{HSB} is the Raman spectra calculated in the Heisenberg model using an effective Raman scattering operator evaluated for $U = 40t$. Ground state transitions as a function of $U$ are highlighted by dashed lines, where the different ground states have been labeled for both the triangular and kagome lattices. } \label{fig:B1g-linecuts}
\end{figure}

Raman light-scattering complements neutron scattering and can be used to understand the nature of spin excitations. Unlike neutron scattering that probes excitations involving odd numbers of spin-flips, Raman scattering enables the probing of even spin-flip excitations in a material that reflects the symmetry of the two dipole transitions compatible with the selected photon polarizations. Formally, Raman scattering can be decomposed into nonresonant, resonant, and mixed responses. Nonresonant scattering, or Thomson scattering, is a two-particle process that directly connects the initial and final states via the stress tensor operator and is only dependent on the energy transfer of the photon, whereas resonant Raman scattering is mediated through excitations to intermediate states via a four particle process~\cite{Tom2007Review}.

In this section, we calculate the resonant Raman scattering cross-section off-resonance using two methods: (1) direct evaluation of the resonant Raman response in the Hubbard model using the Kramers-Heisenberg expression far away from resonance; and (2) an effective Raman response in the Heisenberg model using the Shastry-Shraiman formulation~\cite{ShastryRaman,ShastryRaman_long}, valid when $U$ is much larger than the incident photon energy $\omega_i$ and nearest neighbor hopping $t$. Here, focus is placed on intermediate to strong coupling, where the spin and charge responses are well separated because of the Mott gap. 

We consider the half-filled Hubbard model having charge gaps, and therefore we limit focus on the most resonant diagrams only. We will consider these diagrams when the incident photon energy is tuned away from any direct transition (calling it "off-resonance") in this section, and in the following section we consider nearly "on-resonant" scattering.

In all lattice geometries considered here, the $A_{2g}$ symmetry channel projects to the antisymmetric part of the matrix element where the photon polarization is taken to be $e^f_xe^i_y-e^f_ye^i_x$, effectively probing states that break time-reversal symmetry. In the $A_{2g}$ symmetry channel, the stress tensor cancels exactly, allowing the resonant $A_{2g}$ Raman operator to be expressed as a SSC operator. This expression allows us to investigate the chiral character of strongly correlated system through resonant Raman scattering.

\begin{figure} [!tb]
\begin{center}
\includegraphics[width=1\columnwidth]{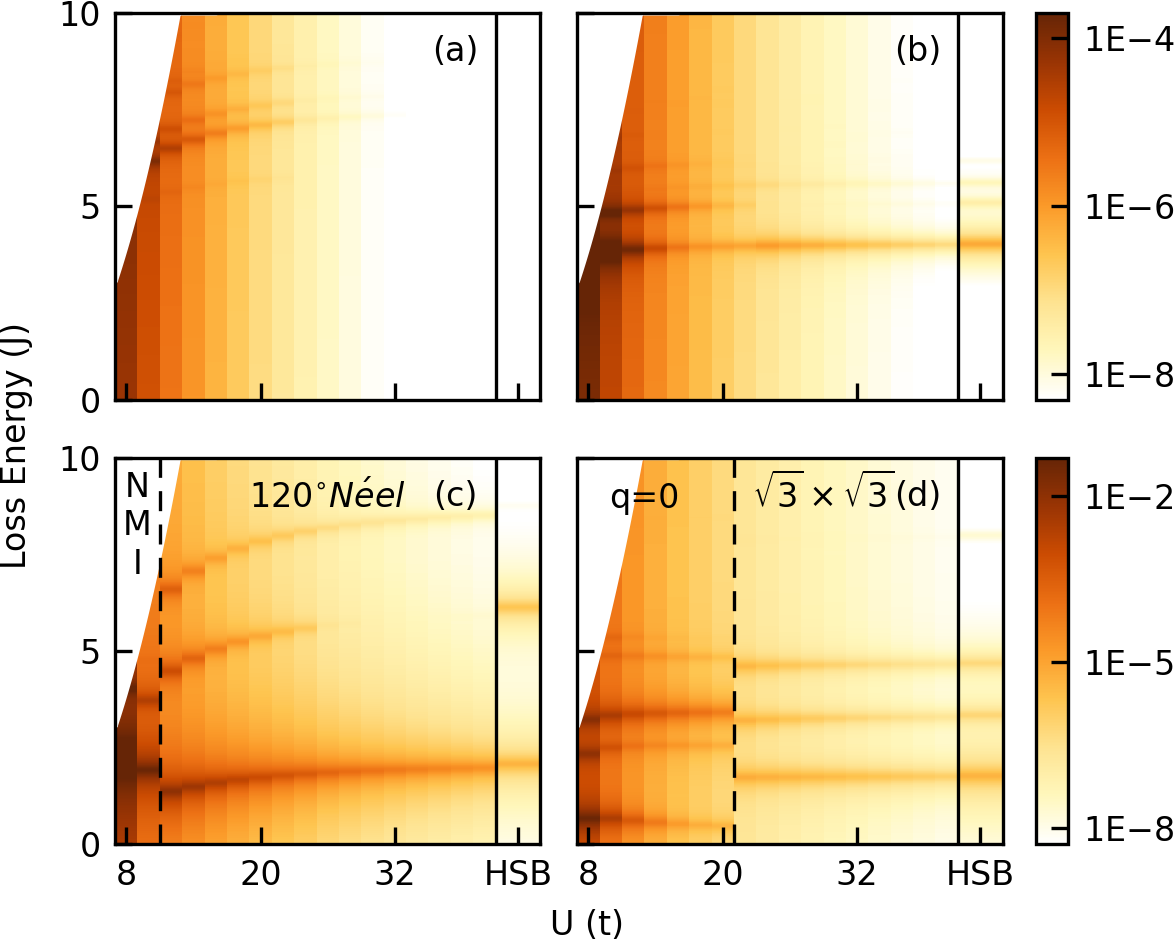}
\end{center}
\caption{\textbf{Off-resonance Raman spectra in $\mathbf{A_{2g}}$ symmetry calculated for  (a)
square lattice (b) honeycomb lattice (c) triangular lattice and (d)
kagome lattices, with incident energy $\mathbf{\omega_i = U/4}$:}  The calculated spectra plotted in this figure have the same layout and use the same broadening parameters as in Figure~\ref{fig:B1g-linecuts}. The intensity for Raman $A_{2g}$ scattering on the square lattice (panel (a)) in the Heisenberg limit ($<$ 1E$-$10) is too small to appear on the plot for this intensity range.}
    \label{fig:A2g-linecuts}
\end{figure}

\subsubsection{Results}

We first study Raman scattering on a two-dimensional square-lattice Hubbard model, where the low-energy physics is closely related to insulating cuprates~\cite{ZhangRice}. Here, we examine the Hubbard model across various coupling strengths while maintaining the same off-resonance condition (where $\omega_i=U/4$) across all values of $U$. In subsequent figures, the exchange energy $J$ is taken as $4t^2/U$, with Lorentzian broadening in incident energy $\Gamma_{in} = 1t$ and in loss energy of $\Gamma_{loss} = 0.1J$ (the same for the Heisenberg model). In all figures, the white region at the left-side of each panel indicates where the loss energy is larger than the incident energy and therefore physically inaccessible.

Figure\figref{fig:B1g-linecuts}{(a)} shows the off-resonant Raman scattering for the square-lattice Hubbard model at half-filling. In the $B_{1g}$ channel, we take the polarizations to be $e^f_xe^i_x-e^f_ye^i_y$ (same as the $E_g$ channel in other lattice geometries), which probes the antiferromagnetic excitations via two-magnon processes~\cite{FleuryLoudon,blumberg1996resonant,chubukov1995resonant,Muschler2010}. Here, the intensity remains strong as the interaction strength increases, displaying a peak at $\sim 3J$, corresponding to the two-magnon excitation energy. In contrast, Fig.\figref{fig:A2g-linecuts}{(a)} shows scattering in the $A_{2g}$ channel, where the polarizations are taken to be $e^f_xe^i_y-e^f_ye^i_x$. The energy loss has a leading peak $\sim 5J$, gradually increasing to $\sim 6J$ at large $U$ and the scattering intensity decays rapidly in this channel, as excited states with chiral character are well separated from the ground state. The rapidly decaying intensity results from high order spin flip processes, with the first non-zero contribution of the effective scattering operator occurring at the order of $t^6/(\omega_i-U)^5$ (see the Supplemental Information~\cite{supplemental}).

For the honeycomb lattice, the two-magnon energy loss, as observed in the $E_{g}$ channel (see Fig.\figref{fig:B1g-linecuts}{b}), is $\sim 1.8J$, corresponding to the energy cost of flipping neighboring spins. On the other hand, the effective Raman scattering operator in the $A_{2g}$ channel is finite at order $t^4/(\omega_i-U)^3$~\cite{KoA2g2010}, leading to stronger scattering intensity in the Raman $A_{2g}$ channel when compared with the square lattice. Indeed, Figure\figref{fig:A2g-linecuts}{(b)} shows that the intensity in the $A_{2g}$ channel decays much more slowly than for the square lattice, and the energy loss is $\sim 4J$ in the intermediate to strong coupling regime ($U>10t$). The gradual increase in charge fluctuations due to reduced coordination in the honeycomb lattice leads to stronger $A_{2g}$ scattering intensity and softening in the energy loss.

On the triangular lattice, frustration can lead to a spectral downshift of the magnon dispersion, detectable in the Raman $E_{g}$ channel~\cite{zheng2006anomalous,vernay2007raman}, as seen in Figure\figref{fig:B1g-linecuts}{(c)}. In the strong coupling limit with 120$^\circ$ N\'{e}el ordering, the leading excitation in the $A_{2g}$ channel occurs $\sim 2J$ (Fig.~\figref{fig:A2g-linecuts}{(c)}), lower than that of both the square and hexagonal lattices, a further indication of the influence of frustration in the system. With decreasing coupling strength, the Raman spectra in the effective Heisenberg model~\cite{tang2022spectra} connects smoothly to the calculated spectra in the Hubbard model. As $U$ changes in the intermediate coupling regime, a ground state transition occurs between $U = 10t$ and $12t$, clearly indicated by a discontinuity in the Raman spectra in both the $E_{g}$ and $A_{2g}$ channels (Figs.~\figref{fig:B1g-linecuts}~and~\figref{fig:A2g-linecuts}{(c)}). 

In Figure\figref{fig:A2g-linecuts}{(d)}, we show that the calculated Raman spectra for the kagome Heisenberg and Hubbard models in the strongly correlated limit have loss energies in the $A_{2g}$ channel $\sim 2J$, similar to the triangular lattice.  While the low-energy excitation is gapped, finite-size studies are needed to ascertain the size dependence of the low-energy excitations observed in the $A_{2g}$ channel as debates remain on the magnetic ordering of the kagome Heisenberg model~\cite{Singh2008kagome_VBC,Ying2007PRL_U1QSL,YanWhite_2011_kagomeQSL}. At the ground state transition where $U\sim20t$, Raman scattering in the $E_g$ (Fig.\figref{fig:B1g-linecuts}{(d)}) (Fig.\figref{fig:A2g-linecuts}{(d)}) channel becomes nearly gapless with a clear discontinuity in the spectral features, indicative of a ground state transition into the $q=0$ state. We emphasize that low-energy excitations in the $A_{2g}$ channel persist in the $q=0$ state, indicating that the ground state of the kagome Hubbard model may lie in close proximity to a chiral spin liquid ground state.

\subsection{Resonant Raman Scattering Near Mott Gap} \label{sec:Res-Raman}

Often, the photon energy used in many Raman scattering experiments may lie near the value of $U$, far from the off-resonance regime. Additionally, some spin liquid candidate materials have been reported to have intermediate values of $U$~\cite{McMahan1990Cuprate,shimizu2003spin}. In both cases, the Shastry-Shraiman approximation made in section~\ref{sec:OffRes-Raman} is no longer valid. Here, we calculate Raman scattering on resonance ($\omega_i = U$) to reveal strong chiral spin excitations that can be accessed when the incident energy is tuned near the upper Hubbard band center. Experimental measurements of cuprates also display enhancing intensity in $A_{2g}$ channel when incident energy is on-resonance~\cite{chelwani2018magnetic}.

\begin{figure} [!tb]
\begin{center}
\includegraphics[width=1\columnwidth]{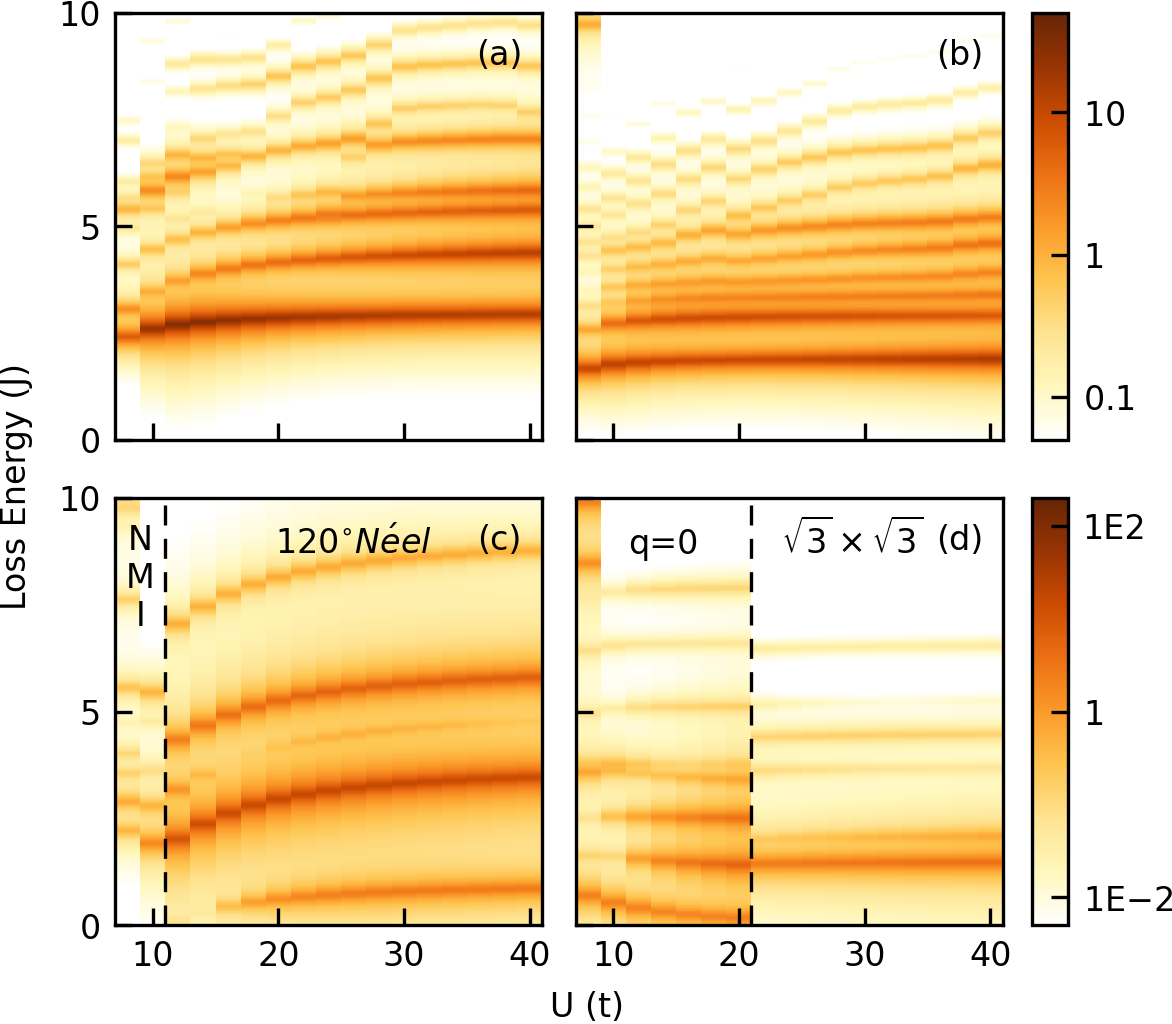}
\end{center}
\caption{\textbf{Resonant Raman spectra in (a)
$\mathbf{B_{1g}}$ square lattice (b) $\mathbf{E_{g}}$ honeycomb lattice (c) $\mathbf{E_{g}}$ triangular lattice and (d)
$\mathbf{E_{g}}$ kagome lattices, with incident energy at $\mathbf{\omega_i = U}$}: The calculated spectra plotted in this figure have the same layout and broadening parameters as in Figure~\ref{fig:B1g-linecuts}. The resonant Raman spectra exhibit a similar excitation profile to the off-resonance Raman spectra in Figure~\figref{fig:B1g-linecuts}{}. However, the peak amplitudes differ from those off-resonance, due to different processes.}
    \label{fig:B1g-linecuts-res}
\end{figure}

Figure\figref{fig:B1g-linecuts-res}{} shows resonant Raman response in the $B_{1g}$ and $E_{g}$ symmetry for each lattice type. For the square and honeycomb lattices, the scattering amplitude is roughly an order of magnitude larger than off-resonance. The scattering intensity no longer decays with increasing $U$; instead, the intensities are on the same order of magnitude across all $U$, as high energy spin excitations become much more visible compared to off-resonance conditions. The overall characteristics of the excitation profile remain similar to that in the off-resonance regime (Fig.~\figref{fig:B1g-linecuts}{}).

For the triangular lattice, the scattering amplitude in the $E_g$ channel is noticeably stronger in the 120$^\circ$ N\'{e}el state than in the NMI state (Fig.\figref{fig:B1g-linecuts}{(c)}). Similarly, the scattering amplitude in the $E_g$ channel is also slightly stronger in the $q=0$ state for the kagome lattice (Fig.\figref{fig:B1g-linecuts}{(d)}). Here, the change in intensity of the two-magnon scattering is consistent with the result shown in the spin structure factor as discussed in section~\ref{sec:Structure-Factor}.

\begin{figure} [!tb]
\begin{center}
\includegraphics[width=1\columnwidth]{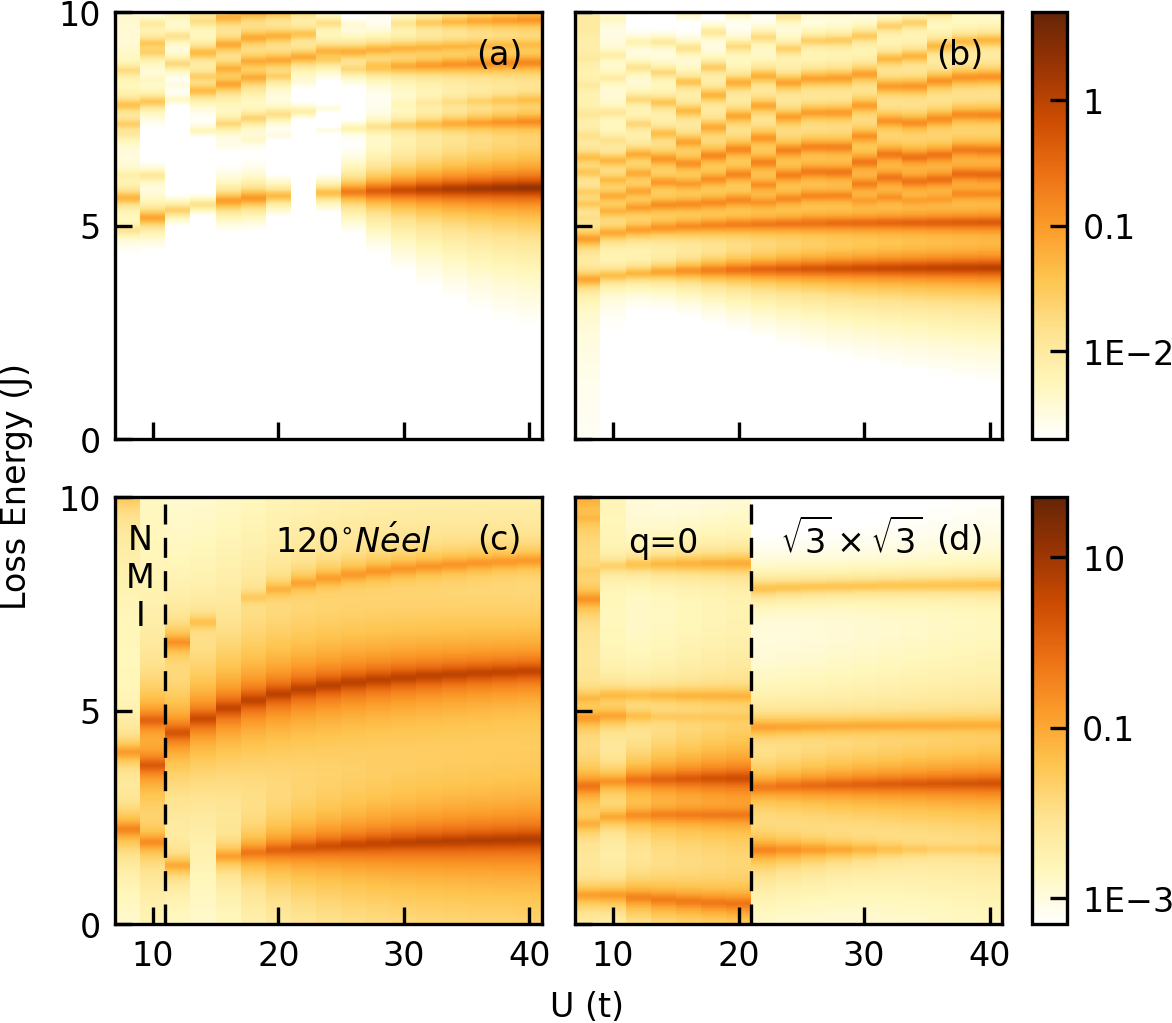}
\end{center}
\caption{\textbf{Resonant Raman spectra in $\mathbf{A_{2g}}$ symmetry calculated for (a)
square lattice (b) honeycomb lattice (c) triangular lattice and (d)
kagome lattices, with incident energy $\mathbf{\omega_i = U}$}: The calculated spectra plotted in this figure have the same layout and broadening parameters as Figure~\ref{fig:B1g-linecuts}, with the same photon polarization as in Figure~\ref{fig:A2g-linecuts}.}
    \label{fig:A2g-linecuts-res}
\end{figure}

Resonant Raman scattering in the $A_{2g}$ symmetry channel shows increasing scattering intensity as a function of $U$ in (Fig.~\figref{fig:A2g-linecuts-res}{}), in stark contrast to the off-resonance scattering, where the intensity decreases exponentially as $U$ increases (Fig.~\figref{fig:A2g-linecuts}{}). On the square lattice, we see an enhancement of chiral spin excitations (Fig.~\figref{fig:A2g-linecuts-res}{(a)}) around 5$J$, previously weak in the off-resonance regime. By tuning the incident energy near resonance, we recover leading chiral spin excitations that are difficult to detect off-resonance. We note that previous experimental Raman measurements of insulating cuprates~\cite{SulewskiA2g1991,Muschler2010} also display enhanced intensity in $A_{2g}$ channel when the incident photon energy is on-resonance. Although we show in Section~\ref{sec:OffRes-Raman} that the Raman scattering intensity in the $A_{2g}$ channel diminishes quickly with the off-resonance condition, resonant Raman scattering can still be an effective probe of chiral spin excitations, amplified by the resonance effect.

A similar enhancement of the $A_{2g}$ scattering intensity can also be seen in honeycomb and triangular lattices, where the scattering intensity is now only an order of magnitude smaller than that of the $E_{g}$ channel across all values of $U$. More notably, the Raman cross section in the $A_{2g}$ channel for the kagome lattice is of the same order of magnitude as the other channels, with low lying chiral spin excitations visible in the $q=0$ state (Fig.~\figref{fig:A2g-linecuts-res}{(d)}), similar to off-resonance Raman scattering (Fig.~\figref{fig:A2g-linecuts}{(d)}). Here, the enhanced scattering intensity in $A_{2g}$ symmetry makes resonant Raman scattering a promising tool for probing the existence of spin liquids with chiral character in materials with a kagome lattice geometry, especially those with large values of $U$, such as in herbertsmithite-like compounds~\cite{Jeschke2013PRB}.

\section{Discussion}

We have analyzed the spin structure factor and resonant Raman cross section for the Hubbard model on different lattice geometries. The spin structure factor shows diminishing characteristics of antiferromagnetism with increasing frustration. Furthermore, ground state ordering can change with decreasing coupling strength in frustrated geometries. Finally, we employed Raman scattering in different scattering channels to understand the change in underlying symmetry of the ground state and the development of chiral character with increased geometric and charge frustration.

In the strongly correlated limit and off-resonance, the Raman scattering operator in the $A_{2g}$ channel can be expressed in terms of scalar spin chirality to reveal the presence of chiral spin excitations. We showed that the strength of chiral spin fluctuations is highly dependent on the lattice geometry. Bipartite lattices, such as the square and honeycomb lattices, have much weaker scattering in the $A_{2g}$ symmetry channel compared to frustrated triangular and kagome lattices. We observed a softening of the energy in the $A_{2g}$ Raman response for both frustrated lattice geometries at the boundary of the respective ground-state transitions, indicating the emergence of chiral character in the intermediate coupling regime, as also shown in large cluster studies~\cite{xu2024hubbard,sun2021metal}. Meanwhile, a clear discontinuity in the Raman spectra can be seen in all symmetry channels across the ground-state transition. For the kagome lattice, we observed a near-gapless excitation in the Raman $A_{2g}$ channel at intermediate coupling strength ($U \sim 20t)$, suggesting that a possible chiral spin liquid phase may be stabilized in the Hubbard model on the kagome lattice with an appropriate perturbation~\cite{Claassen2017dynamical}.

Although the Raman $A_{2g}$ scattering intensity is typically small in the off-resonance regime, it can become comparable in magnitude to other scattering channels when the incident photon energy is tuned near resonance. Figure~\figref{fig:A2g-linecuts-res}{} clearly shows that the Raman $A_{2g}$ scattering intensity increases to the same order of magnitude as the two-magnon response in other scattering channels on-resonance. Low-lying chiral spin excitations are accessible through the resonant processes when $\omega_{in} \sim U$ for all lattice geometries, allowing for reliable detection of chiral character in Mott insulators.

Our paper establishes the utility of Raman $A_{2g}$ scattering for detecting chiral fluctuations in Mott insulators. However, a few open questions remain. Firstly, we note that finite-size effects are large in the small clusters considered in this paper, and we cannot directly capture the proposed spin liquid ground states in the triangular lattice Hubbard model in the intermediate coupling regime~\cite{shirakawa2017ground,szasz2020chiral,chen2022quantum,xu2024hubbard} or the kagome lattice Heisenberg models~\cite{Ying2007PRL_U1QSL,YanWhite_2011_kagomeQSL}. Numerical methods such as density matrix renormalization group (DMRG) allow for much larger cluster sizes, but it remains difficult to evaluate the resonant diagram using those techniques~\cite{NoceraDDMRG2018}. Nevertheless, we emphasize that Raman $A_{2g}$ scattering effectively detects chiral character independent of finite-size effects, and opens an avenue for numerical studies on larger clusters and future experiments. 

Additionally, we have only considered Raman scattering in Hubbard models with nearest-neighbor interactions. However, similar studies may be performed on more sophisticated models that can better describe different exotic phases, such as the inclusion of long-range interactions or antisymmetric exchange terms that introduce additional frustration beyond lattice geometry, which may also lead to the realization of a CSL phase. Without relying on an effective Raman scattering operator, one can compute and study the chiral spin excitations by evaluating the Kramers-Heisenberg formula for resonant Raman scattering in the $A_{2g}$ channel to study the emergence of CSL states in frustrated Mott insulators.

\section*{Acknowledgments}

We thank T. Tang, J.K. Ding and M. Claassen for helpful discussions and suggestions. This paper was supported by the U.S. Department of Energy (DOE), Office of Basic Energy Sciences, Division of Materials Sciences and Engineering. Computation work was performed on the resources of the National Energy Research Scientific Computing Center (NERSC) supported by the U.S. Department of Energy, Office of Science, using NERSC award BES-ERCAP0027203. C. Jia acknolwedges the support from Center for Molecular Magnetic Quantum Materials, an Energy Frontier Research Center funded by the U.S. Department of Energy, Office of Science, Basic Energy Sciences under Award no. DE-SC0019330. 
% \clearpage
\bibliography{main}

\end{document}

%%%%%% Title %%%%%%
\title{Supplementary Information: Detection of chiral spin fluctuations driven by frustration in Mott insulators}
%%%%%% Authors %%%%%%
\author{Kuan H. Hsu}
\affiliation{Department of Materials Science and Engineering, Stanford University, CA 94305, USA}
\affiliation{Stanford Institute for Materials and Energy Sciences,
SLAC National Accelerator Laboratory, 2575 Sand Hill Road, Menlo
Park, CA 94025, USA}

\author{Chunjing Jia}
\affiliation{Department of Physics, University of Florida, FL 32611, USA}

\author{Emily Z. Zhang}
\affiliation{Stanford Institute for Materials and Energy Sciences,
SLAC National Accelerator Laboratory, 2575 Sand Hill Road, Menlo
Park, CA 94025, USA}
\affiliation{Geballe Laboratory for Advanced
Materials, Stanford University, CA 94305, USA}

\author{Daniel Jost}
\affiliation{Stanford Institute for Materials and Energy Sciences,
SLAC National Accelerator Laboratory, 2575 Sand Hill Road, Menlo
Park, CA 94025, USA}

\author{Brian Moritz}
\affiliation{Stanford Institute for Materials and Energy Sciences,
SLAC National Accelerator Laboratory, 2575 Sand Hill Road, Menlo
Park, CA 94025, USA}

\author{Rudi Hackl}
\affiliation{School of Natural Sciences, Technische Universit{\"a}t M{\"u}nchen, Garching 85748, Germany}
\affiliation{IFW Dresden, Helmholtzstrasse 20, Dresden 01069, Germany}

\author{Thomas P. Devereaux}
\email{tpd@stanford.edu}
\affiliation{Department of Materials Science and Engineering, Stanford University, CA 94305, USA}
\affiliation{Stanford Institute for Materials and Energy Sciences,
SLAC National Accelerator Laboratory, 2575 Sand Hill Road, Menlo
Park, CA 94025, USA}
\affiliation{Geballe Laboratory for Advanced
Materials, Stanford University, CA 94305, USA}
\date{\today}

\maketitle

In this Supplementary Information, we present a derivation of the effective Raman scattering operator in $A_{2g}$ symmetry on a square lattice, including the consideration of next-nearest- ($t^\prime$) and next-next-nearest-neighbor hoppings ($t^{\prime\prime}$). 

\section{Derivation of effective Raman scattering operator} \label{A2g-derivation}

Here, we derive the effective Raman scattering operator in $A_{2g}$ symmetry on a square lattice, which has been proposed as a means to probe scalar spin chirality; We can start be treating the Hubbard model coupled to light via the Peierls construction:

\begin{equation} \label{eq:Hubbard-Peierls}
\begin{aligned}
H &= H_{Hubbard} + H_{\gamma} + H_{int} \\
\mathit{H_{int}} &=\frac{e}{ \hbar c} \hat{\mathbf{j}} \cdot \mathbf{A} + \frac{e^2}{2 \hbar^2 c^2} \sum_{\alpha\beta}\mathbf{A}_\alpha \hat{\gamma}_{\alpha\beta} \mathbf{A}_\beta + \dotsm,
\end{aligned}
\end{equation}
where $H_{\gamma} = \omega_{\gamma}n_{\gamma}$ is the free photon Hamiltonian. The expression for $H_{int}$ of the light-matter interaction is obtained from the Peierls substitution, expanded to second order. $\mathbf{A}$ is the vector potential, $\hat{\mathbf{j}}$ is the current operator~\cite{Tom2007Review}, and $\hat{\gamma}$ is the stress tensor that contributes to the nonresonant Raman response. By treating $H_{int}$ as a time-dependent perturbation, and considering second order processes for $\mathbf{A}$ that correspond to one-photon-in/one-photon-out, we can obtain the $T$-matrix for Raman scattering:

\begin{equation} \label{eq:TMatrix}
\begin{aligned}
T_{NR} + T_R = H^{(2)}_{int} + H^{(1)}_{int}\frac{1}{E_i-(H_{Hb}+H_{\gamma})+i\eta}H^{(1)}_{int}.
\end{aligned}
\end{equation}

Next, we follow the prescription by Shastry and Shraiman~\cite{ShastryRaman} by recasting the Fermion operators in terms of spin operators. Expanding the resonant term in Eq.~\ref{eq:TMatrix} in the limit of small $t$:

\begin{equation} \label{Rexpansion}
\begin{aligned}
T_{R} =& H^{(1)}_{int}\frac{1}{E_i-(H_{Hb}+H_{\gamma})+i\eta}H^{(1)}_{int} \\
    =& H^{(1)}_{int}\frac{1}{E_i-(H_{U}+H_{\gamma}+i\eta)}
    \sum_{n=0}^{\infty}(H_t\frac{1}{E_i-(H_{U}+H_{\gamma})+i\eta})^nH^{(1)}_{int}
\end{aligned}
\end{equation}
Neglecting the two-photon part of $H_{int}$, $H_t$ is a sum of fermion operators $H_t = -t\sum_{\sigma} c^{\dagger}_{i\sigma} c_{j\sigma}=-t(c^{\dagger}_ic_j)$. $H_{int}$ is also sums of fermion operators with the additional polarization tensor prefactor: $H_{int} = -it(e\cdot\delta)\sum_{\sigma} c^{\dagger}_{i\sigma} c_{j\sigma}=-it(e\cdot\delta)(c^{\dagger}_ic_j)$. For strong correlations ($U \gg t$), the energy separation between the lower and upper Hubbard bands is on the order of the interaction $U$. This allow us to rewrite the denominator in Eq.~\ref{Rexpansion}, where $(E_i-(H_U+H_{\gamma}))^{-1} = (\omega_i-U)^{-1}$. Thus, the scattering operator can be expressed as a series of virtual electron hops:

\begin{equation} \label{Resfinal}
\begin{aligned}
&T_{R} = (e_f\cdot\delta_f)(e_i\cdot\delta_i)\Bigg[ \Bigg.\sum_{i_1,j_1,i_2,j_2} \frac{-t_1t_2}{\omega_i-U}(c^{\dagger}_{i_2}c_{j_2})(c^{\dagger}_{i_1}c_{j_1})  + \sum_{i_1,j_1,i_2,j_2,i_3,j_3} \frac{t_1t_2t_3}{(\omega_i-U)^2}(c^{\dagger}_{i_3}c_{j_3})(c^{\dagger}_{i_2}c_{j_2})(c^{\dagger}_{i_1}c_{j_1})  + \dotsm \Bigg]. \Bigg.
\end{aligned}
\end{equation}
Here, the first term corresponds to the zeroth order expansion of Eq.~\ref{Rexpansion}, and so on. The last step of this procedure is to convert the chain of fermion operators into spin operators using the following spin identities:

\begin{equation} \label{eq:spin-id}
\begin{aligned}
&c^{\dagger}_{\sigma}c_{\sigma'} = \widetilde{\chi}_{\sigma'\sigma} = \frac{1}{2}\delta_{\sigma',\sigma} + \mathbf{S}\cdot \boldsymbol{\tau}_{\sigma'\sigma}
\\&c_{\sigma}c^{\dagger}_{\sigma'} = \chi_{\sigma\sigma'} = \frac{1}{2}\delta_{\sigma,\sigma'} - \mathbf{S}\cdot \boldsymbol{\tau}_{\sigma\sigma'}
\\& (\mathbf{a}\cdot\boldsymbol{\tau})(\mathbf{b}\cdot\boldsymbol{\tau})=(\mathbf{a}\cdot\mathbf{b})I+i(\mathbf{a}\times\mathbf{b})\cdot\boldsymbol{\tau},
\end{aligned}
\end{equation}
where $\boldsymbol{\tau}$ are Pauli matrices and $\mathbf{S} = \frac{1}{2}c^{\dagger}_{\sigma}\boldsymbol{\tau}_{\sigma\sigma'}c_{\sigma'}$ is the spin operator. Using this formulation, we can determine the contribution to different symmetry channels in a $D_{4h}$ point group by decomposing the Hamiltonian in different symmetries. In the following subsections, we will derive the effective Raman operator in the Heisenberg limit for $A_{2g}$ symmetry on a square lattice. At $n = 0$, the expansion will result in the Elliot-Loudon-Fleury (EFL) Hamiltonian~\cite{FleuryLoudon}. Here, we start at $n = 2$. This is the order where there are four virtual hops in the resonant scattering process.

\subsection{Second Order (n = 2)} \label{2nd-order}

The effective $A_{2g}$ operator was previously derived in terms of spin operators shown to cancel up to order $t^4/(\omega_i-U)^3$ due to edge-canceling virtual hopping paths~\cite{KoA2g2010}. We expand upon this result, showing that the effective operator cancels in this order even with the inclusion $t^\prime$ and $t^{\prime\prime}$. Additional care is needed to account for virtual hops: movement of holons and doublons are possible in high order processes. Each path that involves 4 vertices $v1 \rightarrow v2 \rightarrow v3 \rightarrow v4 \rightarrow v1$ can be written down in 4 different orders the of virtual processes~\cite{KoA2g2010,tang2022spectra}.

We first examine the case with the inclusion of $2^{nd}$-nearest-neighbor hopping. To form a close loop, the virtual hopping process will need at least two nearest-neighbor hops and two $2^{nd}$-nearest-neighbor hops (denoted as $\alpha)$. The hopping process can be systematically enumerated by looking at the span of the 4 hops process: (1) All 4 hops are constrained within a unit cell (2) The hopping process spans 2 unit cell, where the initial step starts in the middle of the unit cells (3) The hopping process spans 2 unit cells, however the initial step starts in the corner of the combined unit cells (Fig.~\figref{fig:4hops}{(a-c)}). In total, there are 144 unique pathways that contribute to the $A_{2g}$ channel. By symmetry, we can reduce these pathways using rotations and reflections (the rotations are by 90 degrees and reflection about $x = y$ and $x = -y$) to 18 unique pathways. The individual contributions for each subset of hopping process to the $A_{2g}$ scattering operator, where $\textbf{S}_i\cdot(\textbf{S}_j\times\textbf{S}_k) = S_{ijk}$, are:

\begin{figure}
     \centering
     \includegraphics[width=0.5\columnwidth]{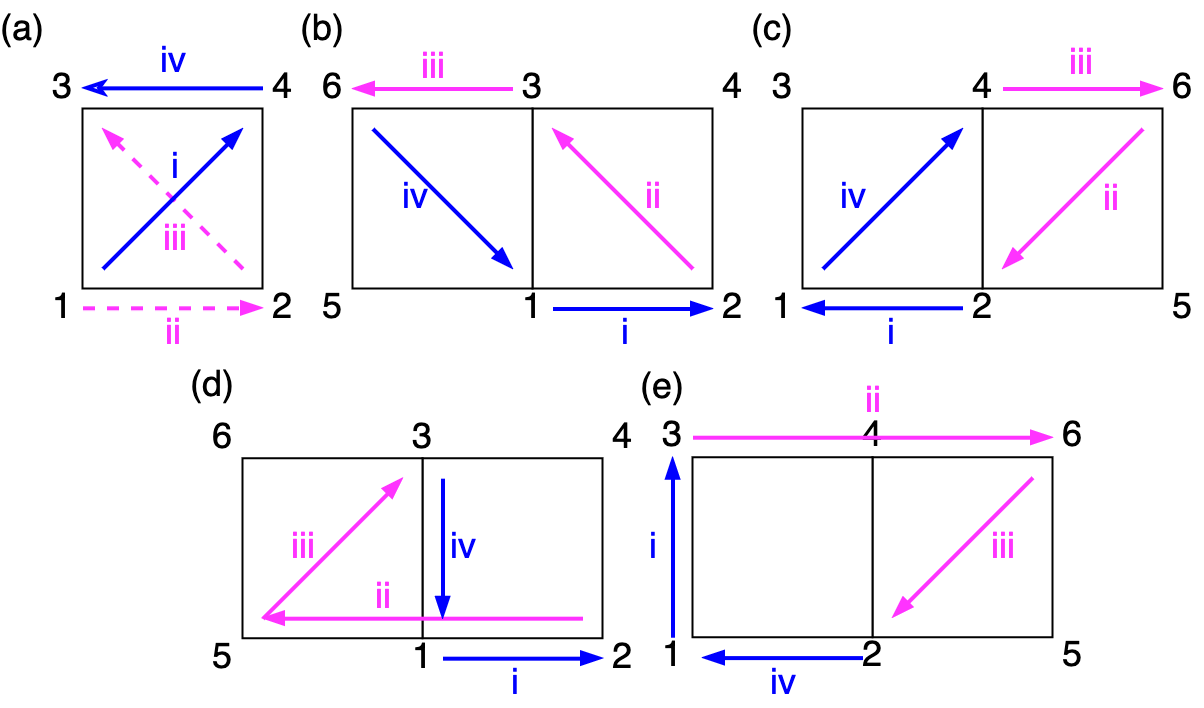}
     \hfill
    \caption{\textbf{Example virtual hopping pathways for second order processes (n=2):} Selected hopping pathways in (a) $O^{\alpha(1)}_{A_{2g}}$ (b) $O^{\alpha(2)}_{A_{2g}}$ (c) $O^{\alpha(3)}_{A_{2g}}$ (d) $O^{\beta(2)}_{A_{2g}}$ (e) $O^{\beta(3)}_{A_{2g}}$. Blue arrow indicates first and final hops of an electron. A solid magenta line represents doublons (one of the electrons from a doubly occupied site moving from site i to create another doubly occupied site at j), while a dashed magenta line represents holons (one of the electrons from a singly occupied site moving from site j to create another empty site at i).}  \label{fig:4hops}
\end{figure}

\begin{equation} \label{eq:2ndorder-op-alpha}
\begin{aligned}
O^{\alpha(1)}_{A_{2g}} &= \frac{t^2{t^\prime}^2}{(\omega_i-U)^3}\sum_{r}8i(S_{r+y,r+x,r}+3S_{r+x+y,r+x,r} 
- Ref. + Rot.) \\&= \frac{it^2{t^\prime}^2}{(\omega_i-U)^3}\sum_{r}32\Bigg( \Bigg.\hspace{4pt}\ta - Ref. + Rot.\Bigg) \Bigg. \\
O^{\alpha(2)}_{A_{2g}} &= \frac{t^2{t^\prime}^2}{(\omega_i-U)^3}\sum_{r}4i(-S_{r+y,r+x,r}+3S_{r+x+y,r+y,r}-S_{r-x,r+y,r} - Ref. + Rot.)
 \\& = -\frac{it^2{t^\prime}^2}{(\omega_i-U)^3}\sum_{r}16\Bigg( \Bigg.\hspace{4pt}\ta - Ref. + Rot.\Bigg) \Bigg. \\
O^{\alpha(3)}_{A_{2g}} &= \frac{t^2{t^\prime}^2}{(\omega_i-U)^3}\sum_{r}16i(-S_{r+x+y,r+x,r} - Ref. + Rot.) %+
\\ & = -\frac{it^2{t^\prime}^2}{(\omega_i-U)^3}\sum_{r}16\Bigg( \Bigg.\hspace{4pt}\ta - Ref. + Rot.\Bigg). \Bigg.
\end{aligned}
\end{equation}
Summing all parts, the resulting expression of $O^\alpha_{A_{2g}} = 0$. We can repeat this process again where there is one $2^{nd}$-nearest-neighbor hop and one $3^{rd}$-nearest-neighbor hop (denoted as $\beta$). Due to the inclusion of $3^{rd}$-nearest-neighbor hop, the hopping pathway must span at least 2 unit cells (Fig.\figref{fig:4hops}{(d,e)}). In total, there are 192 total pathways. The individual contributions for each subset of hopping processes to the $A_{2g}$ scattering operator are:

\begin{equation} \label{eq:2ndorder-op-beta}
\begin{aligned}
\\O^{\beta(2)}_{A_{2g}} = & \frac{t^2t^\prime t^{\prime\prime}}{(\omega_i-U)^3}\sum_{r}i(24S_{r+y,r+x,r}-2S_{r+2x+y,r+x,r}-2S_{r+y,r+2y+x,r} \\& + 6S_{r+x,r+2x+y,r+y}+4S_{r+2y,r+x,r}+4S_{r+y,r+2x,r} + Rot.)
\\ =&  \frac{it^2t^\prime t^{\prime\prime}}{(\omega_i-U)^3}\sum_{r}\Bigg( \Bigg. \hspace{2pt}24\hspace{2pt}\ta -2 \hspace{2pt}\tb -2 \hspace{2pt}\tc + 6\hspace{2pt}\td + 4\hspace{2pt}\te + 4\hspace{2pt}\tf + Rot.\Bigg) \Bigg.
\\O^{\beta(3)}_{A_{2g}} =& -O^{\beta(2)}_{A_{2g}}.
\end{aligned}
\end{equation}
Again, all terms cancel out and $O^{\beta}_{A_{2g}} = 0$. Finally, the case where there are two nearest-neighbor and two $3^{rd}$-nearest-neighbor processes can be neglected, as this case can be easily mapped to the case when there are 4 nearest-neighbor terms and the operators will again cancel. Here, we definitively show that the effective Raman scattering operator in $A_{2g}$ cancels to this order, even with the inclusion of longer range hopping.

\subsection{Third Order (n = 3)}

Similar to the sections above, we can derive an expression for a third order process in terms of spin operators. For a 5 hop pathway, we can always find a cancelling pathway that carries opposite sign of spin chirality terms. Assume a scattering process where the electron moves on the following vertices: $v_1 \rightarrow v_2 \rightarrow v_3 \rightarrow v_4 \rightarrow v_5 \rightarrow v_5$, the operator associated with the pathway is $T_{3}$ from Eq.~\ref{eq:3rdorder}.

\begin{equation} \label{eq:3rdorder}
\begin{aligned}
T_{3} =& C_3(\delta_1,\delta_5)(c^\dag_1c_5)(c^\dag_5c_4)(c^\dag_4c_3)(c^\dag_3c_2)(c^\dag_2c_1) \\ 
=& 2C_3(\delta_1,\delta_5)tr\{\rchi_5\rchi_4\rchi_3\rchi_2\widetilde{\rchi}_1\} \\
\rightarrow& iC_3(\delta_1,\delta_5)(S_{321}+S_{421}+S_{431}-S_{432}-S_{521}+S_{531}-S_{532}+S_{541}-S_{542}+S_{543}).
\end{aligned}
\end{equation}
Here $C_3(\delta_1,\delta_5) = [(e_f^{y}\cdot\delta_5)(e^x_i\cdot\delta_1)-(e_f^{x}\cdot\delta_5)(e_i^y\cdot\delta_1)](t^4{t^\prime})/(\omega_i-U)^4$. Now, consider an alternative process that involves the following order of vertices: $v_2 \rightarrow v_1 \rightarrow v_5 \rightarrow v_4 \rightarrow v_3 \rightarrow v_2$. The operator associated with this pathway can be written down as:

\begin{equation} \label{eq:3rdorder-conj}
\begin{aligned}
T^\ast_{3} =& C_3(-\delta_1,-\delta_5)(c^\dag_5c_1)(c^\dag_4c_5)(c^\dag_3c_4)(c^\dag_2c_3)(c^\dag_1c_2) \\ 
=& C_3(-\delta_1,-\delta_5)tr\{\widetilde{\rchi}_5\widetilde{\rchi}_4\widetilde{\rchi}_3\widetilde{\rchi}_2\rchi_1\} \\
\rightarrow& 2iC_3(-\delta_1,-\delta_5)(-S_{321}-S_{421}-S_{431}+S_{432}-S_{521}-S_{531}+S_{532}-S_{541}+S_{542}+S_{543}) 
\\=& -T_{3}.
\end{aligned}
\end{equation}
We can find canceling pathways for all permutations of the virtual scattering process. Thus, all pathways will cancel and result in no contribution to the $A_{2g}$ scattering operator.

\begin{figure}[!tb]
     \centering
     \includegraphics[width=0.45\columnwidth]{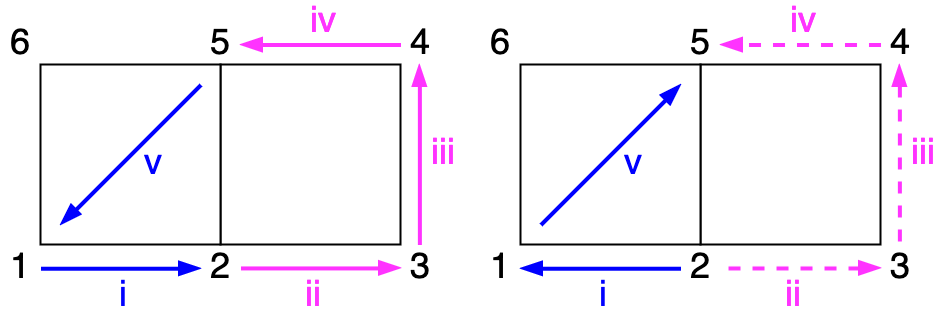}
     \hfill
    \caption{\textbf{Cancelling pathways in third order processes}: As outlined in the text, the two pathways shown above cancel out.}
    \label{fig:5hops}
\end{figure}

\subsection{Fourth Order (n = 4)}

\begin{figure}[!bt]
     \centering
     \includegraphics[width=0.45\columnwidth]{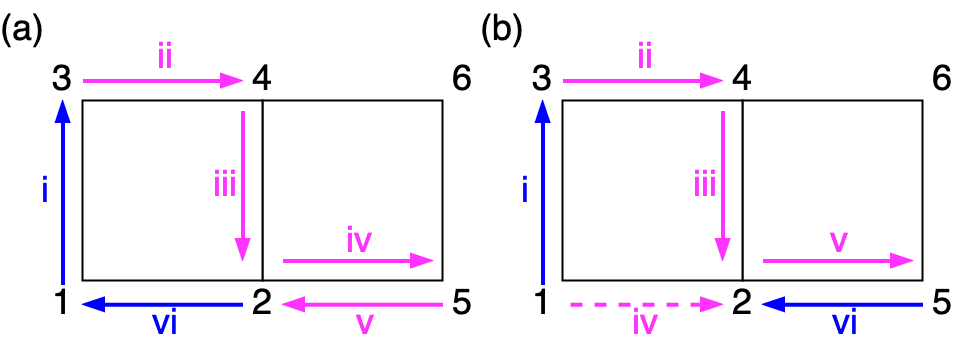}
     \hfill

    \caption{\textbf{Example of fourth order process (n = 4) that is valid (a) and invalid (b):} Both pathways depicted above are permutations of a 6 hop pathway involving vertices: $v_1 \rightarrow v_3 \rightarrow v_4 \rightarrow v_2 \rightarrow v_5 \rightarrow v_2 \rightarrow v_1$. Each permutation has non-collinear first and final hops, resulting in non-zero contribution to Raman response in $A_{2g}$ symmetry. However, in the permutation shown on the right, the holon and doublon recombine on vertex 2, making it an invalid pathway.}
    \label{fig:invalidpaths}
\end{figure}

Continuing the expansion from Eq.~\ref{Rexpansion} to $n = 4$ order, a scattering process with 6 virtual hops with only nearest-neighbor interaction can be written as a chain of fermion operators, and converted to spin operators:

\begin{equation} \label{eq:4thorder-convert}
\begin{aligned}
T_{4,a} =& (e_f\cdot\vec{r}_{f})(e_i\cdot\vec{r}_{i})\frac{t^6}{(\omega_i-U)^5}(c^\dag_1c_6)(c^\dag_6c_5)(c^\dag_5c_4)(c^\dag_4c_3)(c^\dag_3c_2)(c^\dag_2c_1) \\ 
=& C_4(\delta_1,\delta_6)tr\{\rchi_6\rchi_5\rchi_4\rchi_3\rchi_2\widetilde{\rchi}_1\} \\
=&2iC_4(\delta_1,\delta_6)\Bigg( \Bigg. \sum_{6 \geq a > b > c \geq 2}S_{abc} - \sum_{6 \geq a > b \geq 2} S_{ab1}\Bigg). \Bigg.
\end{aligned}
\end{equation}
Here $C_4(\delta_1,\delta_6) = [(e_f^{y}\cdot\delta_6)(e^x_i\cdot\delta_1)-(e_f^{x}\cdot\delta_6)(e_i^y\cdot\delta_1)](t^6)/(\omega_i-U)^5$. There are 16 possible permutations for a pathway that involves 6 vertices. Since the number of pathways and permutations grows significantly with higher order expansions, we have developed a program that counts each pathway and permutation, summing their respective contributions to the $O_{A_{2g}}$ operator. Starting from a random point r, there are 400 unique 6 hop pathways that form a close loop, 360 of which have non-parallel components that contribute to the $A_{2g}$ scattering operator. As before, one needs to examine whether a permutation of a pathway forms a closed loop before the final hop (Fig.\figref{fig:invalidpaths}{(b)}). If so, the pathway can be excluded from consideration. Finally, we can write out the expression for $O_{A_{2g}}$:

\begin{equation} \label{eq:4thorder-operator}
\begin{aligned}
O_{A_{2g}} =& \frac{t^6}{(\omega_i-U)^5}\sum_{r}i(34S_{r+y,r+x,r}+9S_{r+2x+y,r+x,r}+9S_{r+y,r+2y+x,r}
\\&+30S_{r+x,r+2x+y,r+y}+9S_{r+2y,r+x,r}+9S_{r+y,r+2x,r} + Rot.)
\\ =&  \frac{it^6}{(\omega_i-U)^5}\sum_{r}\Bigg( \Bigg. \hspace{2pt}34\hspace{2pt}\ta + 9\hspace{2pt}\tb + 9\hspace{2pt}\tc + 30\hspace{2pt}\td + 9\hspace{2pt}\te + 9\hspace{2pt}\tf + Rotation.\Bigg) \Bigg.
\end{aligned}
\end{equation}

% \clearpage
\bibliography{main}